\documentclass[utf8]{FrontiersinHarvard} % for articles in journals using the Harvard Referencing Style (Author-Date), for Frontiers Reference Styles by Journal: https://zendesk.frontiersin.org/hc/en-us/articles/360017860337-Frontiers-Reference-Styles-by-Journal
\usepackage{url,hyperref,lineno,microtype,subcaption}
\usepackage[onehalfspacing]{setspace}
\usepackage[T1]{fontenc}
\def\keyFont{\fontsize{8}{11}\helveticabold }
\def\firstAuthorLast{Bianchi E.} %use et al only if is more than 1 author
\def\Authors{Eleonora Bianchi\,$^{1,2,*}$}
\def\Address{$^{1}$INAF, Osservatorio Astrofisico di Arcetri, Largo E. Fermi 5, I-50125 Firenze, Italy \\
$^{2}$Univ. Grenoble Alpes, CNRS, IPAG, F-38000 Grenoble, France}
\def\corrAuthor{Eleonora Bianchi}

\def\corrEmail{eleonora.bianchi@inaf.it}

\begin{document}
\onecolumn
\firstpage{1}

\title[iSEEDs]{Astrochemical Study of Early Embedded Disks} 

\author[\firstAuthorLast ]{\Authors} %This field will be automatically populated
\address{} %This field will be automatically populated
\correspondance{} %This field will be automatically populated

\extraAuth{}% If there are more than 1 corresponding author, comment this line and uncomment the next one.
%\extraAuth{corresponding Author2 \\ Laboratory X2, Institute X2, Department X2, Organization X2, Street X2, City X2 , State XX2 (only USA, Canada and Australia), Zip Code2, X2 Country X2, email2@uni2.edu}

\maketitle

\begin{abstract}
The question of how our planet was formed and, more generally, how a planetary system forms is fundamental and has been addressed in a broad range of research domains. However, we still lack a comprehensive understanding of the basic aspects of the process of star and planet formation. 
In particular, the challenge of measuring the mass and chemical composition of young protostellar disks has, so far, hampered a meaningful comparison with observed exoplanet populations. This will become critical in the near future to interpret the results of European space missions, such as \href{https://arielmission.space/}{Ariel}, which will yield a comprehensive inventory of exoplanetary masses and chemical compositions.
Building on recent developments in astrochemistry and data science, this perspective explores future research avenues for the study of young planet-forming disks and introduces the project "Astrochemical Study of Early Embedded Disks" (\href{https://www.iseeds.inaf.it/home}{iSEEDs}).
By integrating machine learning and data mining with astrochemistry, iSEEDs provides a robust framework to systematically extract the physical conditions and molecular abundances hidden within high-resolution datasets of protostellar environments.

%%% Leave the Abstract empty if your article does not require one, please see the Summary Table for full details.

\tiny
 \keyFont{ \section{Keywords:} Astrochemistry, stars and planets formation, interstellar medium, planetary systems, observations and modelling} 
 %All article types: you may provide up to 8 keywords; at least 5 are mandatory.
\end{abstract}

\section{Introduction}

Planet formation occurs as a direct consequence of star formation around a broad range of stellar types \citep[e.g.,][and references therein]{Draz2023}. In particular, the formation of a star similar to our Sun begins when a dense
region within a molecular cloud becomes gravitationally unstable, leading to the creation of a disk due to angular momentum
conservation (\citealt{Shu1987, Palla2004, Bergin2007, Andre2014, Caselli2012, Tobin2024}; see Section \ref{subsec:form-a-cloud}). The disk mediates the accretion of material from the envelope onto the star through accretion streamers, as well as the ejection of material via jets and outflows \citep{Pineda2023, Frank2014, Lesur2023}. Most importantly, the disk is the birthplace of planets: within the disk gas and dust agglomerate and grow forming the protoplanets and their atmospheres, in a process lasting $\sim$ 10 Myr \citep{Testi2014, Draz2023, Birnstiel2024}.
Despite a well-established framework,
numerous fundamental questions persist. Recent evidence suggests that planet formation begins earlier than previously thought, in young disks of less than 0.5 million years around embedded protostars \citep[e.g.,][]{ALMA2015, Sheehan2018, Segura-Cox2020, Maureira2024}.
Observational facilities operating across a range of wavelengths, including the Karl G. Jansky Very Large Array (\href{https://science.nrao.edu/facilities/vla}{VLA}), the Northern Extended Millimeter Array (\href{https://iram-institute.org/observatories/noema/}{NOEMA}), the Atacama Large Millimeter/submillimeter Array (\href{https://www.almaobservatory.org/en/home/}{ALMA}), and the James Webb Space Telescope (\href{https://www.esa.int/Science_Exploration/Space_Science/Webb}{JWST}), offer insights into the
properties of young disks. However, bridging the gap between the physical and chemical properties of protostellar disks and those of the planets they form remains a challenge. The initial conditions for planet formation, which influence the structure and chemical composition of the emerging planetary system, as well as their migration history, are established within their natal disks. Understanding the physical properties and evolution of young embedded disks is therefore crucial, yet disentangling their various physical components, such as the disk, infalling envelope, outflows, and accretion streamers, remains a formidable challenge \citep[e.g.][]{Bianchi2017-HH212, Bianchi2022, Bianchi2023-faraday, Tobin2024}. While CO isotopologues are standard tracers for mature disks, they often prove insufficient in younger, embedded stages due to severe cloud contamination and high optical depth \citep[e.g.,][]{Miotello2023}.

In this context, the "Astrochemical Study of Early Embedded Disks" (\href{https://www.iseeds.inaf.it/home}{iSEEDs}) project was designed to look beyond traditional tracers. By inferring the detailed physical and chemical structure of young planet-forming disks, iSEEDs seeks to determine the actual mass available for planet formation and characterise the initial chemical budget from which planetary systems form. Furthermore, the project aims to pinpoint when and how dust grains begin their growth toward planetesimals within the disk midplane, marking the very first steps of planet building.
Astrochemistry is currently experiencing a period of rapid expansion, marked by significant advancements in our understanding of the processes responsible for molecular formation and, ultimately, the emergence of life \citep{Caselli2012, Ceccarelli2023}. The advent of powerful interferometers has created vast public archives that remain, largely unexploited. While individual observations are typically designed for specific goals, the resulting datasets often harbor a substantial volume of unanalysed data, particularly regarding complex molecular lines, which require specialized expertise to be fully decoded. 
By integrating data mining and machine learning with astrochemistry, the iSEEDs project seeks to characterise the physical and chemical structure of young embedded disks, the birthplaces of planets by simultaneously integrating the analysis of all the observational tracers in telescope archives.
iSEEDs results will have significant implications for diverse fields, including star and planet formation, astrochemistry, exoplanets, and even extragalactic star formation.

\subsection{From a cloud to a planetary system}\label{subsec:form-a-cloud}

\begin{figure}[h!]
\begin{center}
\includegraphics[width=14cm]{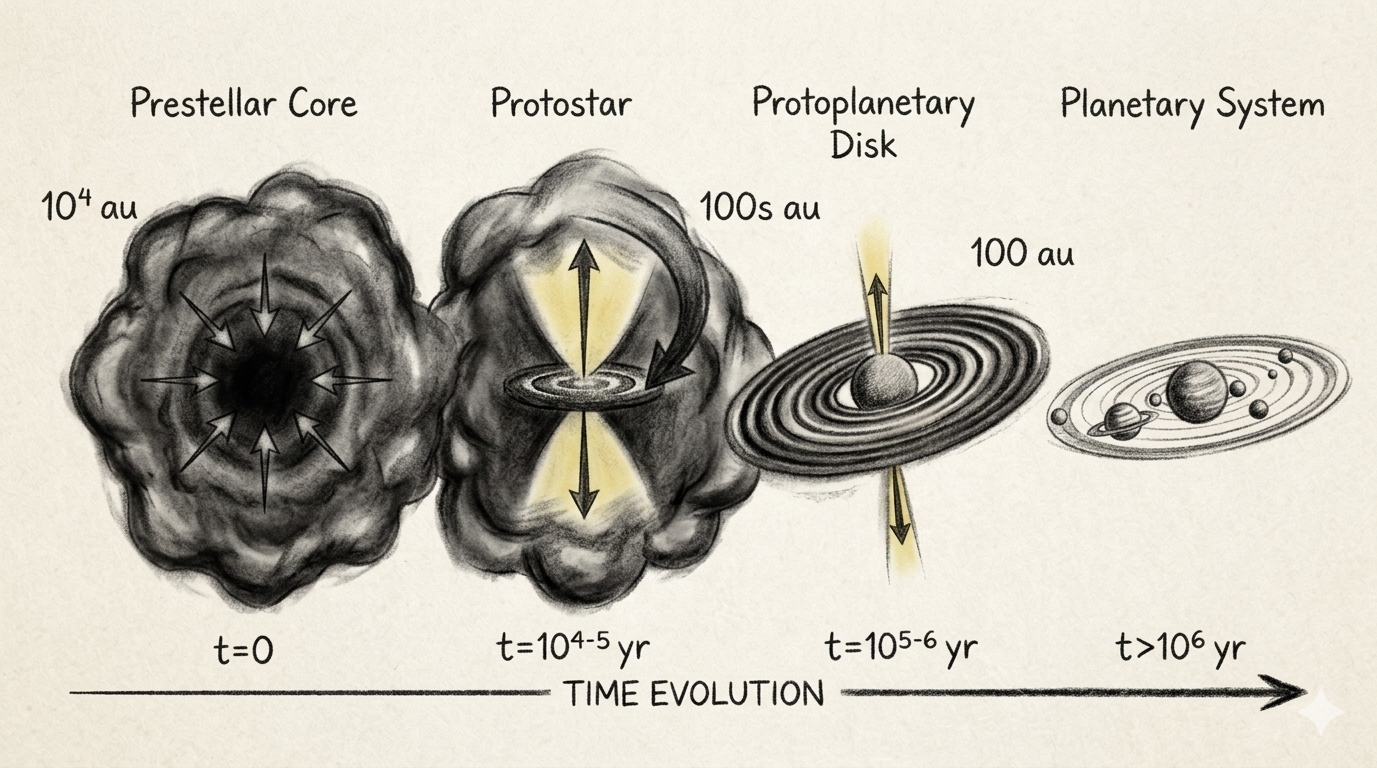}% This is a *.eps file
\end{center}
\caption{Schematic of a Sun-like star and planet formation, showing the transition from a prestellar core to a planetary system. This figure was generated using Gemini (Google) based on an original conceptual sketch provided by the author. }\label{fig:1}
\end{figure}

A major milestone in modern astrophysics is the discovery of more than six
thousand confirmed extrasolar planets, often very different from those of the Solar System \href{https://exoplanet.eu/}{(exoplanet.eu, accessed June 2026)}. In order to understand the origin of such diversity, it is imperative to bear in mind that planets represent a by-product of the star formation process \citep[e.g.,][]{Testi2014, Draz2023, Morbidelli2024}.
In the classical picture, the process of star and planet formation begins when a cold ($\sim$10 K) and dense ($>$10$^4$ cm$^{-3}$) cloud experiences gravitational instability and starts to collapse under its own gravity \citep[e.g.,][]{Shu1987,Palla2004}. 
During the collapse, the material accreting from the large-scale envelope to the central protostellar embryo, conserves its angular momentum and naturally forms a protostellar disk \citep[e.g.,][]{Bate2018, Draz2023, Lebreuilly2024}.
In parallel, a part of the infalling material is ejected by the system and forms collimated bipolar jets and large-scale outflows \citep[e.g.,][]{Frank2014}. Initially, the source is deeply embedded in its parental envelope (Class 0/I sources, 10$^4$-10$^5$ years, \citealt{Lada1987, Andre1993}), which is
gradually accreted onto the star or dispersed by the jet (Class I/II sources, 10$^5$-10$^6$ years), until the newly formed star and its disk remain naked (Class III sources, 10$^7$ yrs, \citealt{Greene1994}).
The disk region, spanning a few hundred astronomical units (au), is the site of planet formation. Over a period of roughly 10 Myr, gas and dust within the disk evolve from simple aggregates into protoplanets, which progressively reshape their environment by carving gaps and cavities until a planetary system is formed \citep[e.g.,][]{Testi2014, Draz2023, Birnstiel2024}.
The transition from a prestellar core to a planetary system, including the characteristic spatial scales and timescales involved, is summarized in the schematic overview of Figure \ref{fig:1}.

\subsection{Young planet-forming disks}\label{subsec:young-disks}
In recent years, the advent of the ALMA interferometer has fundamentally challenged established frameworks of planet formation. More specifically, substructures, composed of rings and gaps, were observed in the dust emission of young disks with estimated ages of a few 10$^5$ years \citep{Sheehan2018, Segura-Cox2020, Maureira2024}. Although their origin is
still debated, it is possible that they are carved by protoplanets. This suggests that planet formation starts early in embedded disks and not in Class II disks with ages of $\sim$ 1 Myr, as previously thought \citep{Testi2014}. This simple outcome raises several issues and open
questions. First of all, if young embedded Class 0/I disks (age $<$ 0.5 Myr) provide the initial conditions of planet formation, their characterisation is crucial in order to make any meaningful comparison with models of planet formation.
Characterising young embedded disks presents various challenges:
\begin{itemize}
    \item[-] The primary challenge lies in the observational
aspect, as young disks are deeply embedded within their natal envelopes. Isolating disk emission from the envelope requires observing complex species and isotopologues with optically thin lines \citep[e.g.,][]{Miotello2023}.

\item[-] Continuum emission is commonly used to derive fundamental disk parameters such as sizes and mass (e.g. CALYPSO
survey \citep{Maury2019}; VANDAM survey \citealt{Segura-Cox2018}) but in young disks a substantial fraction of this emission arises from optically thick regions \citep{Miotello2023, Tazzari2021, Ohashi2023, Maureira2026}, and thus the solid mass budget is systematically underestimated \citep{Manara2018, Tychoniec2020}.

\item[-] During these initial stages disks experience the impact of higher mass accretion rates and variability \citep{Fisher2023}. These processes can deeply change the disk structure and affect how planets form at these early stages.
\end{itemize}  
In summary, while observational capabilities have advanced significantly, fundamental properties such as the size, mass, and temperature of early embedded disks remain largely unconstrained (see \citealt{Miotello2023} for a review). Characterizing their physical and chemical evolution is essential not only to decode the initial stages of planet formation but also to establish a clear link with the diversity of observed exoplanet populations \citep{Miotello2023, Pacetti2022, Pacetti2025}.\\

\begin{figure}[h!]
\begin{center}
\includegraphics[width=14cm]{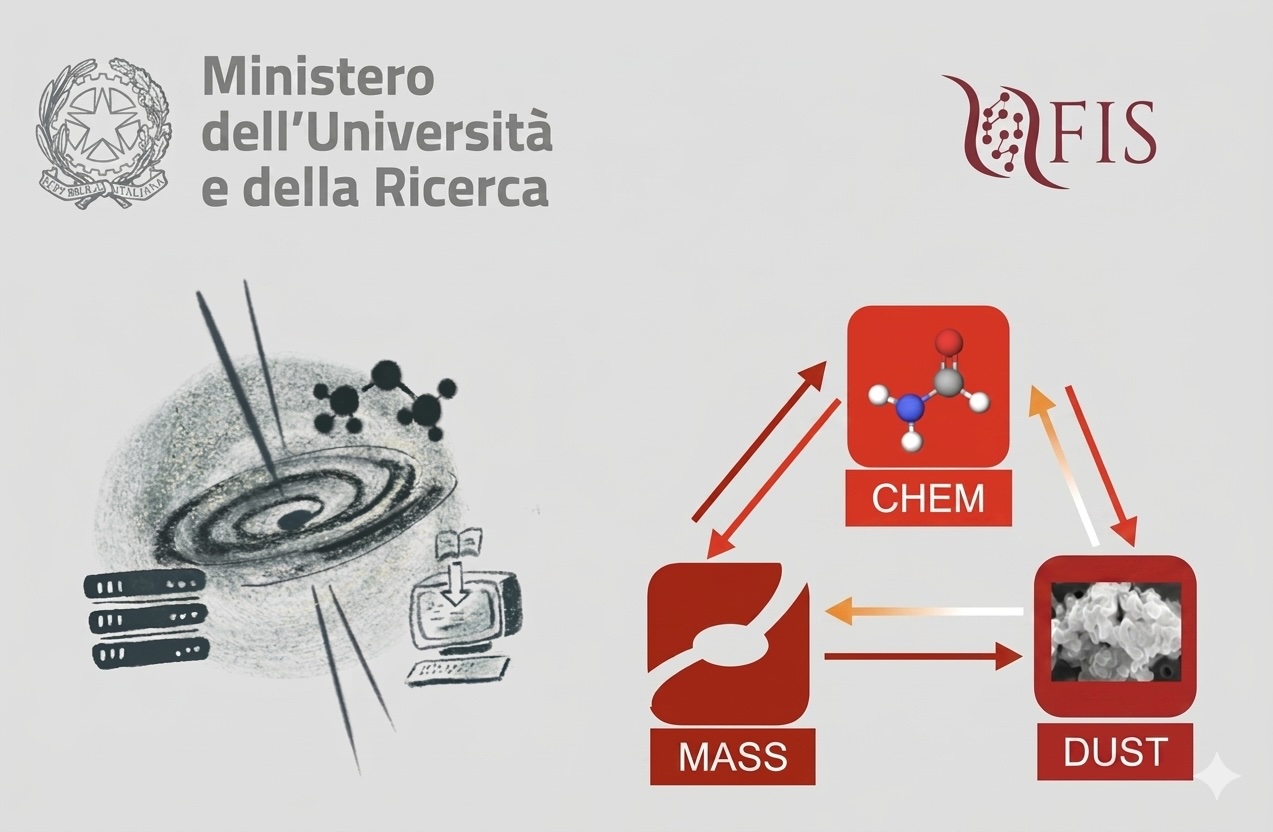}% This is a *.eps file
\end{center}
\caption{The "Astrochemical Study of Early Embedded Disks" (\href{https://www.iseeds.inaf.it/home}{iSEEDs}) project employs an interdisciplinary approach, integrating astrochemistry and data science to characterise the mass, chemical composition, and dust properties of a statistically significant sample of young planet-forming disks. Image credits: logo from the Italian Ministry of University and Research; project illustration designed by \href{https://linktr.ee/gianilivia}{L. Giani}; scientific scheme by the author. Composition assembled using Gemini (Google).}\label{fig:2}
\end{figure}

\section{The Astrochemical Study of Early Embedded Disks (iSEEDs) project}\label{sec:iSEEDs}
The iSEEDs project is a 36-months research project funded by the Italian Ministry of University and Research (\href{https://fis-submission.mur.gov.it/}{MUR}) through the Italian Science Fund (\href{https://fis-submission.mur.gov.it/}{FIS}).
It is designed to bridge the current gaps in our understanding of the physical and chemical properties of disks during their earliest stages. 
By moving beyond single-object studies toward a statistically significant sample, the project is positioned to provide a more comprehensive framework for addressing three fundamental questions:
\begin{itemize}
    \item [1.] How much mass is available in the disk for planet formation?
    \item [2.] What is the chemical composition of disks, possibly inherited by forming planets?
    \item [3.] When and how do dust grains grow in the disk midplane to form planetesimals?
\end{itemize}
iSEEDs will use an innovative interdisciplinary approach, combining astrochemistry, data mining and machine learning techniques.
Complex molecular species, such as interstellar complex organic molecules (molecules with more than 5 atoms and containing Carbon, Oxygen and/or Nitrogen \citep{Ceccarelli2023, Herbst2009}, deuterated species (molecules where deuterium replace one or more hydrogen atoms; \citealt{Ceccarelli2014}) and complex carbon
chains and rings \citep{McGuire2020, Cernicharo2021}, represent the building blocks of prebiotic compounds and it is very relevant to determine if they are at least partially inherited by forming planets. However, this is not the sole point of interest.
The efficient exploitation of a wide range of molecular lines, in many cases already available in open access public archives, can be used to selectively investigate the physical processes down to the scales where protoplanets are forming. While the interpretation of complex molecules relies on a close synergy with chemists, these species have demonstrated to be a powerful tool for the exploration of different physical processes taking place in young embedded disks, such as accretion streamers, accretion shocks, and disk atmospheres \citep{Bianchi2023-faraday, Garufi2022, Lee2017-atmospheres}. 
High-resolution data of complex molecular lines remain largely unexploited within the public archives of major interferometric facilities like ALMA and NOEMA.
By applying advanced data mining and machine learning, we can move beyond individual case studies toward a statistically robust census of planet-forming environments across all currently observed systems.\\
iSEEDs brings together a team of researchers with mixed expertise in observational astrochemistry and data science, fostering a cross-disciplinary environment to extract maximum physical insight from telescopes archival data. This interdisciplinary approach is structured around the three main research lines illustrated in Figure \ref{fig:2}: 
\begin{itemize}
    \item[1.] quantifying the disk mass and the contribution of accreted material from streamers;
     \item[2.] characterising the disk’s chemical composition along with the vertical and radial gas distribution, and 
      \item[3.] unveiling the properties of the dust component.
\end{itemize}
By combining the results from these three interconnected research lines, iSEEDs provides the necessary framework to unveil the key physical and chemical properties of planet-forming systems and their evolution.

\subsection{MASS. The building of the disk and the role of accretion streamers}
Standard planet formation models predict that a disk cannot convert 100\% of its material into planets, meaning exoplanetary system masses should be lower than observed disk masses.
 On the contrary, we find that exoplanetary systems' masses are comparable or higher than the most massive disks \citep{Manara2018,Tychoniec2020}. This evident problem of missing mass can arise from multiple reasons. First, as explained in Section \ref{subsec:young-disks}, disk masses and radii derived from (sub-)mm continuum emission are systematically underestimated due to high optical depth.
When deriving gas mass, H$_2$ is not detectable in sub-mm observations due to its lack of a dipole moment. As a result, CO, the second most abundant molecule after H$_2$, is often utilised \citep{Miotello2016}. However, in young Class 0/I disks, emission lines from CO and its isotopologue, $^{13}$CO, are frequently optically thick. Even in the case of the rarer isotopologue C$^{18}$O, if it is not optically thick, distinguishing the emission from the disk, the outflow, and the envelope can still pose significant challenges. The key to addressing this issue lies in astrochemistry, as employing multiple molecular tracers, which are less abundant ($\lesssim$10$^{-8}$ with respect to H$_2$), enables to disentangle the disk emission \citep[e.g.,][]{Bianchi2017-HH212, Codella2018, Lee2022}. If we observe multiple rotational transitions from a species for which collisional coefficients are available, we can use non-local thermodynamic equilibrium (non-LTE) large velocity gradient (LVG) analysis \citep[see e.g.,][]{Ceccarelli2003, Bianchi2020, Bianchi2022, Bianchi2017-SVS13} to place meaningful limits on the local H$_2$ gas density.
Promising tracers for young disks are formaldehyde (H$_2$CO) and its isotopologues, methyl cyanide (CH$_3$CN) and its isotopologues, methanol (CH$_3$OH) and its isotopologue ($^{13}$CH$_3$OH, CH$_3$$^{18}$OH).

In addition to the mass already included within the disk, estimates of the total mass budget need to consider the impact of high mass accretion rates and variability \citep{Fisher2023}.
Theoretical studies and magnetohydrodynamic (MHD) simulations predicted that accretion of gas from cores with an initially perturbed velocity distribution, occurs through streamers of material funnelled into the disk \citep{Kratter2010, Kuffmeier2019, Hennebelle2020, Calcino2025}. Accretion streamers have been only recently detected in observations of protostellar systems \citep[e.g.,][]{Alves2019, Pineda2020, Pineda2023, Hsieh2023, Valdivia2024, Codella2024, Tanious2024}. These structures connect the large-scale envelope ($\sim$10$^2$ to $\sim$10$^4$ au), to the protostellar disk (scales $<$ 100 au), causing in some cases shocks at the impact zone \citep{Garufi2022,Bianchi2023-faraday, Podio2024}.
Accretion streamers replenish protostellar disks from their natal environment at rates estimated between 10$^{-5}$--10$^{-6}$ M$_{\odot}$ yr$^{-1}$ \citep{Valdivia2022, Pineda2023, Tanious2025}. This means that streamer infall rates could even exceed protostellar accretion rates derived from NIR spectroscopy.
Quantifying how frequent accretion streamers are and how much mass they provide to young disk is of paramount importance to address the problem of the missing mass.
Furthermore, the streaming material experience shocks at the centrifugal barrier, the
transition zone between the envelope and the disk, or directly at the impact point with the disk \citep{Oya2016, Bianchi2023-faraday, Garufi2022, Sakai2014, Podio2024}. These shocks can
alter both the physical disk structure with a local increase of the gas temperature and the disk chemical composition, having an impact on the first steps of planet formation \citep{vanGelder2021}. 
\\
A primary objective of the iSEEDs MASS research line is to identify candidate accretion streamers and evaluate their occurrence rate. The project is designed to estimate the mass of both streamers and disks by combining machine learning classification with kinematic and radiative transfer modeling. This approach, utilising multiline LTE and non-LTE analyses in conjunction with findings from the DUST research line, seeks to facilitate a systematic comparison between observational data and theoretical models of star and planet formation \citep{Kuffmeier2019, Lebreuilly2021}.

\subsection{CHEM. The chemical composition of young disks}\label{subsec:chem}

The chemical investigation  of the inner regions of young, low-mass protostars has revealed a wide variety of chemical compositions. 
Hot corinos are regions of about 100 au around Solar-type protostars heated up to temperatures larger than 100 K, where molecules are thermally sublimated from the icy mantles of dust grains \citep{Ceccarelli2000, Cazaux2003, Bottinelli2004}. They are are enriched in interstellar complex organic molecules (iCOMs \citealt{Herbst2009,Ceccarelli2023}). Examples of iCOMs include methyl formate (HCOOCH$_3$), ethanol (CH$_3$CH$_2$OH), formamide (NH$_2$CHO) and glycolaldehyde (HCOCH$_2$OH). These molecules are released from the icy mantles of dust grains in the dense ($>$ 10$^7$ cm$^{-3}$) inner 100 au region around the protostar, which is heated up to temperatures larger than the molecules evaporation temperature ($\gtrsim$100 K). The temperature increase causes the sublimation of icy mantles formed on dust grains during the prestellar phase, enriching the gas with molecules that were previously trapped in the ice. This release of molecules into the gas triggers further chemical reactions, potentially leading to the formation of additional iCOMs (see \citealt{Ceccarelli2023} for a review).
Due to their small sizes and the associated observational challenge, the origin of the observed molecular emission in hot corinos is still debated. The release of the observed complex organic molecules may occur in the inner protostellar envelope \citep{Ceccarelli2004} or in the disk atmospheres, due to thermal evaporation of the
icy mantles \citep{Lee2017-atmospheres}, or in the accretion shocks, due to the grain sputtering \citep{Vastel2022,Bianchi2023-faraday}. Only a few objects have been investigated at high enough angular resolution to resolve iCOMs emission on scales of tens of au \citep{Bianchi2022, Lee2022, Maureira2022, Okoda2022, Bianchi2023-faraday, Frediani2025}, thus allowing to investigate their physical origin.
Hot corinos are not ubiquitous among protostars, and a chemically distinct type of protostar also exists. Warm Carbon-Chain Chemistry (WCCC) objects \citep{Sakai2013, Taniguchi2024-WCCC} notable abundance of unsaturated hydrocarbons, carbon chains, and rings, such as C$_4$H, C$_4$H$_2$, and c-C$_3$H$_2$. The best example is the protostar L1527 which was extensively studied and found to lack iCOM emission \citep{Sakai2008, Sakai2010-L1527}. Since then, a few other protostars with similar chemistry have been identified \citep{Sakai2009, Imai2016}. The existence of hybrid sources where the two different chemistry can coexist on different scales, a lukewarm region rich in carbon chains surrounding a very compact hot corino enriched in iCOMs, further complicates the picture \citep{Imai2016, Oya2017, Imai2016, Okoda2023}.\\
From a chemical point of view, the diversity observed among protostars is determined by the composition of the ice mantles that form on dust grains during the cold phase preceding protostar formation. An ice mantle enriched in methane (CH$_4$) leads to WCCC when it sublimates at around 25 K, whereas an ice mantle rich in methanol (CH$_3$OH), sublimating at approximately 100 K, leads to hot corino chemistry \citep{Sakai2013, Ceccarelli2023}. The key question that remains unresolved is what causes the variation in the composition of the ice on dust grains. The initial explanation suggested that the duration of the cold prestellar phase plays a crucial role: shorter timescales ($\lesssim$10$^5$ yr) would favor CH$_4$ ice enrichment and WCCC, while longer timescales ($\sim$10$^6$ yr) would lead to significant CO formation and depletion, resulting in the production of CH$_3$OH and successively to hot corino chemistry after methanol sublimation. However, this is still controversial and chemical models have challenged this view \citep{Aikawa2020}. Various environmental factors present during the prestellar stage, such as density, temperature, and exposure to ultraviolet (UV) or cosmic rays, have been proposed as alternative influences shaping the molecular composition of the future protostar. For instance, external UV irradiation could break CO molecules, freeing carbon atoms and promoting the formation of carbon chains \citep{Spezzano2017, Bianchi2023-gbt}.  Recent surveys have focused on studying the occurrence of hot corino and WCCC sources across different star-forming regions. Specifically, the Perseus ALMA Chemistry Survey (PEACHES) observed 50 protostars in the Perseus star-forming complex, identifying hot corinos in 56\% of them, which represents a substantial proportion \citep{Yang2021}. The ORion Alma New GEneration Survey (ORANGES), another survey completely analogous to PEACHES in terms of sensitivity, spatial resolution, and spectral setup, was conducted targeting 19 protostars in the Orion Molecular Cloud (OMC). Unlike in Perseus, the results revealed that hot corinos are less prevalent in Orion, with an occurrence rate of 26\% \citep{Bouvier2021, Bouvier2022}. This suggests that the environment plays a crucial role in shaping the chemistry of young protostars and, potentially, the planetary systems they form.
Further insights on this will emerge from the large program FAUST (Fifty AU Study of the chemistry in the disk/envelope system of Solar-like protostars), the first ALMA Large Program focused on astrochemistry, designed to survey the chemical composition of 13 young protostellar regions, from large envelope scales ($\sim$2000 au) down to planet-formation scales ($\sim$50 au) \citep{Codella2021, Bianchi2020}.\\
While the chemical composition of young planet-forming disk is not fully clarified, the chemistry of accretion streamers which feeds them is largely unexplored \citep{Pineda2023}. They are detected in dust and gas, in a variety of molecular tracers: dust continuum \citep{Cacciapuoti2024}, scattered light \citep{Ginski2021}, CO \citep{Ginski2021, Huang2022, Gupta2023}, C$^{18}$O \citep{Valdivia2022,Flores2023, Kido2023}, HCO$^{+}$ \citep{Yen2019, Garufi2022}, H$_2$CO \citep{Valdivia2022,Valdivia2023}, SO \citep{Aso2023, Flores2023,Codella2024, Tychoniec2024}, CS \citep{Garufi2022}, CCS \citep{Pineda2020}, CN \citep{Tychoniec2024}, DCN \citep{Hsieh2023}, HNC \citep{Murillo2022}, HC$_3$N \citep{Pineda2020, Murillo2022, Taniguchi2024-streamer}, HC$_5$N \citep{Pineda2020, Murillo2022, Taniguchi2024-streamer}. In only few cases, the gas temperature has been determined using a multi-line analyses \citep{Codella2024, Podio2024}. So far, no studies have been undertaken to assess their overall chemical composition. Streamers transport "chemically fresh" material from the envelope, which differs from the material that has already been processed and frozen into the disk. As a result, the chemical composition of the streamer material may differ from that of the disk. Interestingly, in the disk of Oph16 IRS 63, an enhancement of deuterated molecules has been observed at the impact point of the streamer \citep{Podio2024} and in the disk of SVS13A an accretion shock have been detected using deuterated water (HDO) \citep{Bianchi2023-faraday}. If the material transported by the streamer is incorporated into the disk, it will be, at least partially, inherited by the forming planets. Forming planets can further contribute to modify the disk composition at later stages leaving chemical signatures due to shocks or the irradiation of material within the carved cavities \citep{Facchini2021, Rampinelli2024, Rampinelli2025}.
Understanding the chemical composition of the streamer-disk system at early stages is essential for constraining the initial chemical composition of disks and successively predict the chemical makeup of resulting planets \citep[e.g.,][]{Pacetti2025}.

The iSEEDs CHEM research line is dedicated to characterise the full chemical composition and radial and vertical distribution of molecular species within protostellar disks and their accretion streamers. By using simultaneous observations of selected species as probes of distinct physical environments and comparing these data with state-of-the-art astrochemical models, the project seeks to verify the chemical differentiation on planet-formation scales and to clarify the physical origin of the observed emission. Chemical modeling will be performed using in-house and publicly available codes, such as GRAINOBLE \citep{Taquet2012}, UCLCHEM \citep{Holdship2017}, and RADMC-3D \citep{Dullemond2012}, to investigate molecular formation and destruction routes, and to determine whether these species are released into the gas phase via thermal sublimation or accretion shocks.
The close collaboration with experts in laboratory experiments and quantum chemistry ensures the use of the most current chemical networks and molecular parameters \citep{Ferrero2020, Tinacci2023, Giani2023, Giani2025,Kakkar2025}.
The investigation of radial and vertical profiles is designed to clarify if the majority of complex molecular species sublimate into the gas phase at the water snowline, coinciding with the sublimation of the bulk of icy dust grain mantles, or if individual molecules possess distinct snow-surfaces, potentially distributed over a broader temperature range \citep{Tinacci2023-snowline,Bariosco2025,Boitard2025,Boitard2026}. Modelling the disk structure will further elucidate the implications of these differences for the planet-formation process. Finally, the analysis of radial profiles aims to identify potential new tracers to establish a chemical connection with exoplanets \citep{Miotello2023, Pacetti2025}.

\subsection{DUST. The role of the dust}\label{subsec:dust}
Understanding when and how dust begins to grow into planetesimals is a fundamental step in the study of planet formation \citep[e.g.,][]{Testi2014,Draz2023, Birnstiel2024}.
Evidence of grain growth has been reported in Class II protoplanetary disks, as well as in Class 0/I protostellar envelopes \citep[e.g.,][]{Kwon2009, Miotello2014, Cacciapuoti2025} and, more recently, within outflow cavity walls \citep{Sabatini2024, Sabatini2025}.
Despite these findings, early-stage grain growth in Class 0/I sources has only been observed at the protostellar disk scale in a limited number of cases. Recent studies of individual sources suggest that dust grains may already begin to grow during the Class I \citep{Harsono2018, Aso2025, Han2023} and Class 0 stages \citep{Zamponi2025, Radley2025}. The use of multi-wavelength observations has proven essential, given the high optical depth of young disks at these early stages \citep{Ohashi2023, Maureira2022, Maureira2024}.

Beyond characterising the dust itself, such a multi-frequency approach is critical for the correct interpretation of molecular emission. Because the inner regions of young protostars and the disk mid-plane are characterised by severe continuum obscuration \citep{Carrasco2019, Lee2017-dustlane, Maureira2026}, molecular lines can be partially or entirely shielded. This effect compromises the derivation of fundamental gas properties, such as temperature and column density \citep{DeSimone2020, Bianchi2022}. A striking example of this challenge was observed in the NGC 1333 IRAS 4A binary system. While ALMA millimeter-wave observations were limited by this dust shielding \citep{Lopez-Sepulcre2017}, centimeter-wave VLA data successfully unveiled methanol emission that was previously hidden \citep{DeSimone2020}.
A multi-wavelength approach is thus required to evaluate the dust optical depth in the inner regions of young disks to correctly interpret the observed molecular morphology and chemical segregation within specific sources or binary systems \citep{Bianchi2022, DeSimone2020}. Ultimately, these steps allow for the correction of measured molecular abundances before comparing them on a consistent physical basis.
The characterisation of the dust properties and their impact on molecular line data constitutes the primary focus of the DUST research line.

\section{iSEEDs methodology: Data Science for astrochemistry}

The challenge of characterising the physics and chemistry of young, planet-forming disks requires a fundamental shift in methodology. Consequently, interdisciplinarity forms the foundation of the iSEEDs research strategy. By employing multiple gas and dust tracers, astrochemistry allows for the determination of the mass, temperature structure, and chemical composition of young disks across various evolutionary stages. On the other hand, data science enables a transition from an object-by-object approach to a comprehensive analysis of the entire observed population, a shift currently hindered by traditional methods that require significant manual intervention.
While machine learning has been utilized in astronomy for over three decades \citep[e.g.,][]{Odewahn1992, Miller1996, Ball2006}, its application in astrochemistry is still in its infancy \citep[e.g.,][]{Holdship2018, Grassi2022, Lee2021-ML,Shay2025, Verm2025}. Data mining and data science techniques have great potential to enhance the efficiency of both the selection of young protostellar disks from telescope archives and the subsequent analysis of their molecular line emission. Recent advancements highlight this potential; for instance, self-supervised contrastive representation learning, have been developed to generate summary views of morphologically similar images within the ALMA archive \citep{Stoehr2025}. A similar approach can be employed to classify candidate disks and their accretion streamers in archival observations.
Beyond classification, data science techniques can be implemented to perform precise automated line identification and preliminary radiative transfer analysis. Invertible neural networks have already been successfully deployed to extract stellar parameters and classify spectra in the \href{https://www.gaia-eso.eu/}{Gaia-ESO} survey \citep{Kang2023, Candebat2024}, while convolutional neural networks have demonstrated the ability to automatically identify twenty molecular species in line-rich hot core observations \citep{Kessler2025}. A promising path forward lies in leveraging homogeneous, publicly accessible data from observational \textbf{large programs} as a template for developing and validating automated pipelines. The iSEEDs project aims to spearhead this transition by scaling these methods across the extensive public archives of telescopes such as ALMA. This approach enables the move beyond individual case studies toward the investigation of statistically significant populations, which is essential for a robust understanding of disk evolution. Ultimately, the development of these automated tools will provide a framework for efficient analysis, allowing the broader community to more effectively exploit the vast wealth of existing observatory resources.

\section{Conclusions and future}
It is becoming increasingly clear that a systematic characterisation of young disks is essential for establishing the initial conditions of planet formation, capturing the moment when planetary seeds begin to emerge in analogs of the Solar System precursor. Systematic measurements of key observational parameters, such as disk mass, radius, and chemical composition, across various evolutionary stages provide the vital empirical inputs needed for planet population synthesis models. These models serve as a critical tool to discriminate among competing formation mechanisms by transforming the initial conditions of a disk into a predicted distribution of planetary architectures.
Specifically, by incorporating observed chemical gradients, using for example elemental ratios such as C/O, as key diagnostics, into models, we can predict the resulting atmospheric compositions of planetary populations \citep[e.g.,][]{Pacetti2025}. Furthermore, the integration of new quantum chemical results on molecular binding energies allows us to more accurately predict ice line locations and their direct impact on the chemical makeup of both exoplanets and our own Solar System \citep{Tinacci2023-snowline, Boitard2025, Boitard2026}.
In this context, iSEEDs, together with other community efforts, will be instrumental in providing the necessary constraints from the disk stage, enabling a robust interpretation of the rich data expected from the \href{https://arielmission.space/}{Ariel} mission and the Extremely Large Telescope (\href{https://elt.eso.org/}{ELT}.) 
Ultimately, this effort will enable the community to relate observed planetary chemical signatures back to the specific physical and chemical environments of their formation.
This scientific roadmap is inherently tied to a necessary evolution in data methodology. As the volume of astronomical data grows, programs like iSEEDs and other international initiatives are highlighting that the future of the field depends on the efficient extraction of information from telescope archives. This is both an environmental and technical imperative; with the introduction of a new generation of facilities, such as the Square Kilometre Array Observatory (\href{https://www.skao.int/en}{SKAO}) and its projected 700 petabytes of annual data, the long-term storage of raw data is becoming unsustainable. Consequently, the community is beginning to develop advanced tools for automated processing and analysis within telescope archives.
Ultimately, the synergy of chemical diagnostics and advanced computational methodologies provides the framework necessary to transform observational data into a comprehensive understanding of how a protoplanetary disk form and evolve, establishing the causal link between the environment of birth and the nature of the resulting worlds.
% For Original Research articles, please note that the Material and Methods section can be placed in any of the following ways: before Results, before Discussion or after Discussion.

%%Figures, tables, and images will be published under a Creative Commons CC-BY licence and permission must be obtained for use of copyrighted material from other sources (including re-published/adapted/modified/partial figures and images from the internet). It is the responsibility of the authors to acquire the licenses, to follow any citation instructions requested by third-party rights holders, and cover any supplementary charges.

\section*{Conflict of Interest Statement}
%All financial, commercial or other relationships that might be perceived by the academic community as representing a potential conflict of interest must be disclosed. If no such relationship exists, authors will be asked to confirm the following statement: 

The authors declare that the research was conducted in the absence of any commercial or financial relationships that could be construed as a potential conflict of interest.

\section*{Author Contributions}
The author confirms being the sole contributor of this article and has approved it for publication.

\section*{Acknowledgments}
E.B. acknowledges support from the Italian Ministry for Universities and Research under the Italian Science Fund (FIS 2 Call – Ministerial Decree No. 1236 of 2023 August 1) grant FIS-2023-00170.
The author acknowledges the use of Gemini 3 Flash (Paid Tier), Google, for English language polishing, to generate Figure \ref{fig:1} based on a conceptual sketch provided by the author and to assemble Figure \ref{fig:2}.

%\section*{Data Availability Statement}
%The datasets [GENERATED/ANALYZED] for this study can be found in the [NAME OF REPOSITORY] [LINK].
% Please see the availability of data guidelines for more information, at https://www.frontiersin.org/about/author-guidelines#AvailabilityofData

\bibliographystyle{Frontiers-Harvard} %  Many Frontiers journals use the Harvard referencing system (Author-date), to find the style and resources for the journal you are submitting to: https://zendesk.frontiersin.org/hc/en-us/articles/360017860337-Frontiers-Reference-Styles-by-Journal. For Humanities and Social Sciences articles please include page numbers in the in-text citations 
\bibliography{Mybib}

%%% Make sure to upload the bib file along with the tex file and PDF
%%% Please see the test.bib file for some examples of references

%\section*{Figure captions}

%%% Please be aware that for original research articles we only permit a combined number of 15 figures and tables, one figure with multiple subfigures will count as only one figure.
%%% Use this if adding the figures directly in the mansucript, if so, please remember to also upload the files when submitting your article
%%% There is no need for adding the file termination, as long as you indicate where the file is saved. In the examples below the files (logo1.eps and logos.eps) are in the Frontiers LaTeX folder
%%% If using *.tif files convert them to .jpg or .png
%%%  NB logo1.eps is required in the path in order to correctly compile front page header %%%

%%% If you don't add the figures in the LaTeX files, please upload them when submitting the article.
%%% Frontiers will add the figures at the end of the provisional pdf automatically
%%% The use of LaTeX coding to draw Diagrams/Figures/Structures should be avoided. They should be external callouts including graphics.

\end{document}